\documentclass[prb,twocolumn,floatfix]{revtex4}

\usepackage{graphicx}

\usepackage{dcolumn}

\begin{document}

\title{Thermal Roughening and Deroughening at Polymer Interfaces in \\
Electrophoretic Deposition}

\author{\bf Frank W.\ Bentrem\footnote{To whom correspondence should be addressed.}}
\affiliation{\it Marine Geosciences Division, Naval Research Laboratory, Stennis Space Center, MS 39529}
\author{\bf Ras B.\ Pandey}
\affiliation{\it Department of Physics and Astronomy, The University of 
Southern Mississippi,
Hattiesburg, MS 39406-5046}

\date{\it February 2005}

\begin{abstract}

Thermal scaling and relaxation of the interface width in an electrophoretic deposition of
polymer chains is examined by
a three-dimensional Monte Carlo simulation on a discrete lattice. Variation of the equilibrium interface width 
$W_r$ with the temperature $T$ shows deroughening $W_r \propto
T^{-\delta}$, with 
$\delta \sim 1/4$, at low temperatures and roughening $W_r \propto
T^{\nu}$, with
$\nu \sim 0.4$ at high temperatures. The roughening-deroughening transition temperature $T_t$ increases with longer chain lengths and is
reduced by using the slower segmental dynamics.

\end{abstract}

\maketitle

\section{Introduction}

Polymer chains and
particulates of all shapes and sizes form deposits when they are
driven towards a substrate.
A continuous release and deposition of driven polymer chains may 
lead to a polymeric material with a well defined density at the substrate, 
a growing bulk, and an evolving interface \cite{wool}. 
The characteristics of polymer conformation and density at the substrate, 
bulk, and the interface depend on factors such as the nature of the segmental
dynamics of the polymer chains, polymer-substrate and polymer-polymer
interactions and their entanglement, molecular weight, temperature,
type of medium, etc. Similar issues emerge if the polymer
chains are driven by an electric field as in gel electrophoresis
\cite{andrews81,hoagland96,perkins97} and deposit
on the pore boundary of the gel. Another example would be a surface coating
resulting from polymer chains deposited under a pressure gradient such as
in spin-coating. The applications of such electrophoretic (driven) deposition 
and fabrication processes are enormous with a vast list of 
challenging questions in a relatively large parameter space \cite{van_der_biest99}. 
Using a discrete-lattice computer simulation model, 
we would like to focus here on the basic issue of how the
interface roughness depends on temperature. 

We know that, as the polymer chains deposit on an impenetrable substrate, the polymer density increases and the interface develops
\cite{fp1, fp2, bpf, pandey04}. 
The fluctuation in the height (measured from the substrate) of the polymer coating along the longitudinal (i.e., growth) 
direction is the interface width $W$ and is a measure of the surface
roughness. The interface width
is found to grow in time with power laws
($W = At^{\beta_1}$, $Bt^{\beta_2}$), where $\beta_1$ is
the initial growth exponent followed by $\beta_2$) \cite{fp2,bpf,family,barabasi,halpin-healy95,bxp}
before saturating to a steady-state (constant) value
$W_s$ in the asymptotic regime as the polymer chains continue to
deposit. For the interface growth of particles (this could be viewed as the
extreme limit of reducing the chain length) deposition \cite{family,barabasi,halpin-healy95},
the saturation of the interface occurs as the height-height correlation 
length exceeds the substrate length. For the electrophoretic 
deposition of polymer, such a relation between the height-height
correlation length and substrate length is less clear \cite{fp2,bpf}. 
The problems become
much more complex due to additional relevent lengths such as radius of gyration,
complex density-density correlations (intra- and inter-chain), and their
dependence on temperature, field, etc. In fact, we have
observed interesting roughening and deroughening of the interface
as various influences (driving mechanisms such as temperature and
field \cite{bxp})
compete and cooperate. 

Recently it was found \cite{bxp} that the steady-state interface width 
$W_s$ may
reduce to a lower value ($W_r$) if the system is relaxed by continuing segmental movements of existing chains without
adding new chains into the system. We have noted contrast in 
the characteristics of the relaxed interface width $W_r$ including its dependence
on the molecular weight from that of the steady-state width $W_s$. 
In our previous study \cite{bxp}, we examined the effect of the driving field
on the interface width. Dependence of the interface width and morphology on the
temperature for particle deposition has been recently observed experimentally
\cite{setzu98,stoldt00,caspersen01,zhang04} as well as with Monte Carlo simulations
\cite{stoldt00,caspersen01,kalke01}. In fact, the variation of the interface width
with temperature is shown, under certain conditions, to be nonmonotonic
\cite{stoldt00,kalke01}. Polymer surfaces are also affected by the
deposition temperature \cite{bxp,lee00,pandey04}. In this article, we address the dependence of the interface width on the temperature using a variety of chain lengths, sample sizes and segmental dynamics
after relaxing the polymer chains by ``turning off'' the source of new chains. In the following,
we will briefly introduce
the simulation model and follow with a discussion of the results and conclusions.

\section{Method}
We use our previous computer simulation model \cite{bpf,bxp,bxp2} on a discrete cubic lattice of size
$100 \times L \times L$ with {$L = 4$--40.} This is a
coarse-grained model with the charged polymer (polyelectrolyte) chains represented by a number of contiguous occupied lattice sites, each assigned a unit charge. The chains are initially placed near one end of the sample (near $x=1$) by a
self-avoiding walk and are then driven toward the substrate/wall at the
opposite end ($x = 100$) by a field $E$. The length of a chain $L_c$ is the number of lattice sites it occupies. For the simulations reported here,
$L_c=10$--200. In addition to excluded
volume we consider 
nearest-neighbor and polymer-polymer repulsive interactions as well as polymer-wall attractive 
interactions. Two different combinations of segmental chain dynamics \cite{binder}, 
kink-jump and crankshaft (KC) (slower) \cite{verdier62} and kink-jump, crankshaft, and reptation (KCR) (faster) 
are used to move the chains according to the Metropolis algorithm. The energy change for a chain segment due to the field is 
$$\Delta U=E\Delta x,$$ where $\Delta x$ is the displacement parallel to the field. The repulsive inter- and intra-chain (Coulombic) interaction and attractive polymer-substrate interaction is represented by the energy potential 
\begin{equation}
U = J\sum_{ij} \rho_i \rho_j, 
\label{eq:potential}
\end{equation}
where $J=1$, $\rho_i = 1$ for every occupied lattice site $i$, $\rho_i = 
0$ for all unoccupied sites, and $\rho_i = -1$ for every site along the substrate ($x=101$). The summation in eq~\ref{eq:potential} is over all nearest-neighbor sites in the lattice. An open boundary condition
is used along the longitudinal ($x$) direction and a periodic boundary 
condition along the transverse ($y$,$z$) directions. An attempt to move each node
once is defined as one Monte Carlo Step (MCS). One polymer chain is released into the sample every $32L_c/L^2$ MCS for a constant deposition rate of 0.0005 monomer units per MCS per substrate lattice site. After 75,000 MCS, we allow the sample to relax by stopping the release of new chains but continuing to drive the remaining chains with the electric field until equilibrium is reached. The simulations are performed
for many independent samples to calculate the average values of the physical
quantities, i.e., density profile, interface width, radius of gyration, etc.
Further, we have used different sample sizes to check for the finite size 
effects. Most of the data presented here were generated on a $100 \times 40 
\times 40$ lattice with 10-20 independent simulations, and we have not noticed
severe finite size effects on the qualitative nature of the physical
quantities, except for low temperatures ($T < 0.3$). The unit for all length quantities is the lattice constant, 
and arbitrary units are used for the temperature and electric field.

\begin{figure}
\centerline{\includegraphics*[width=8.25cm]{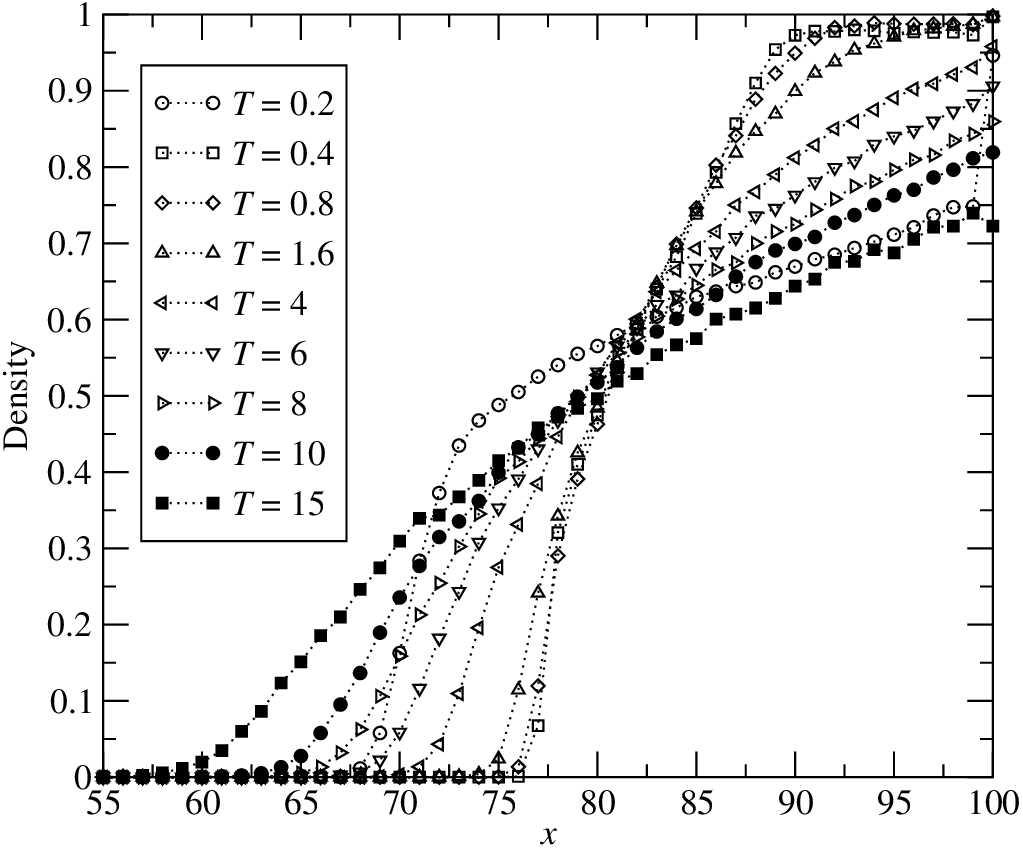}}
\caption{Density profile using a combination of kink-jump and
crankshaft (KC) segmental dynamics. The sample size is $100 \times 40
\times 40$ with $L_c = 40$ and $E = 0.5$ at various temperatures (in
arbitrary units) for 10--20 independent
samples. Density is the fraction of occupied lattice sites and $x$ has
units of the lattice constant.}
\label{fig:density}
\end{figure}

\section{Results}

Figure~\ref{fig:density} shows the equilibrium density profile with the KC dynamics at 
different temperatures with $E=0.5$ and $L_c=40$. Note that the shape of the density profile depends on
the temperature. It is worth pointing out that we have also carried out our 
simulations at lower temperature (i.e., $T < 0.1$) with KC dynamics, 
but encountered technical problems, namely clogging, before deposition and 
a long relaxation time. 
On the other hand, the interface width (along with all other physical 
quantities such as density profile, radius of gyration, etc.) has
reached equilibrium at all temperatures we have studied with the fastest (KCR)
dynamics. The shape of the density profiles using KCR dynamics is 
similar to those in Figure~\ref{fig:density}.

\begin{figure}
\centerline{\includegraphics*[width=8.25cm]{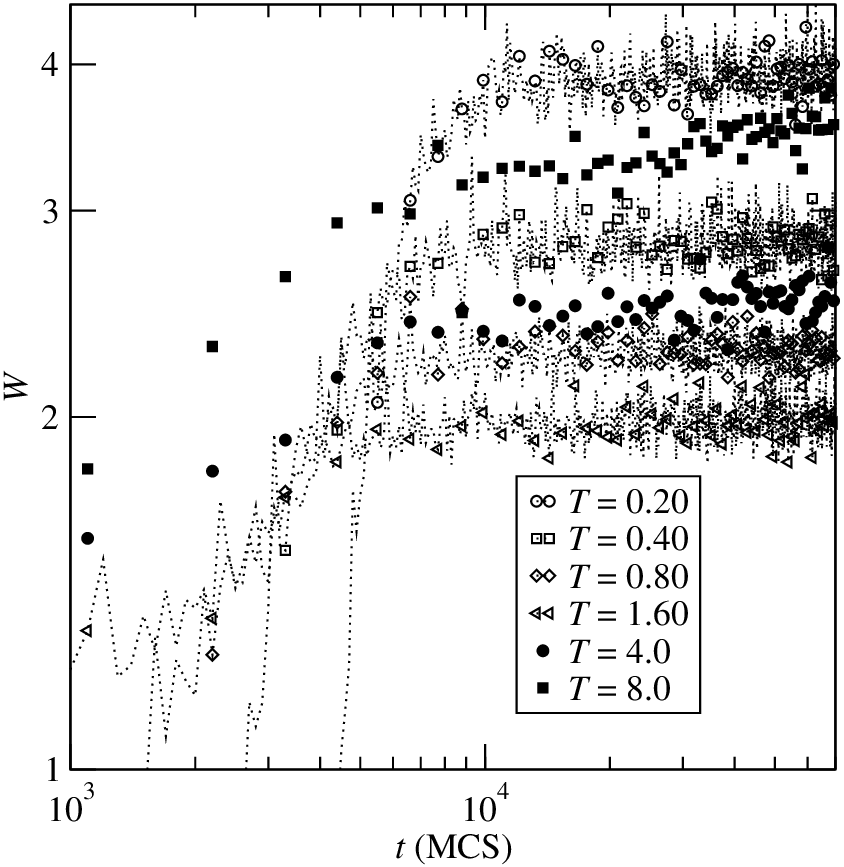}}
\caption{Interface width $W$ (in units of the lattice constant) versus time steps $t$ before surface relaxation on 
a log-log scale using KCR segmental
dynamics for selected temperatures (in arbitrary units) and 10--20 independent samples.
Other parameters are the same as in Figure\ 1.}
\end{figure}

\begin{figure}
\centerline{\includegraphics*[width=8.25cm]{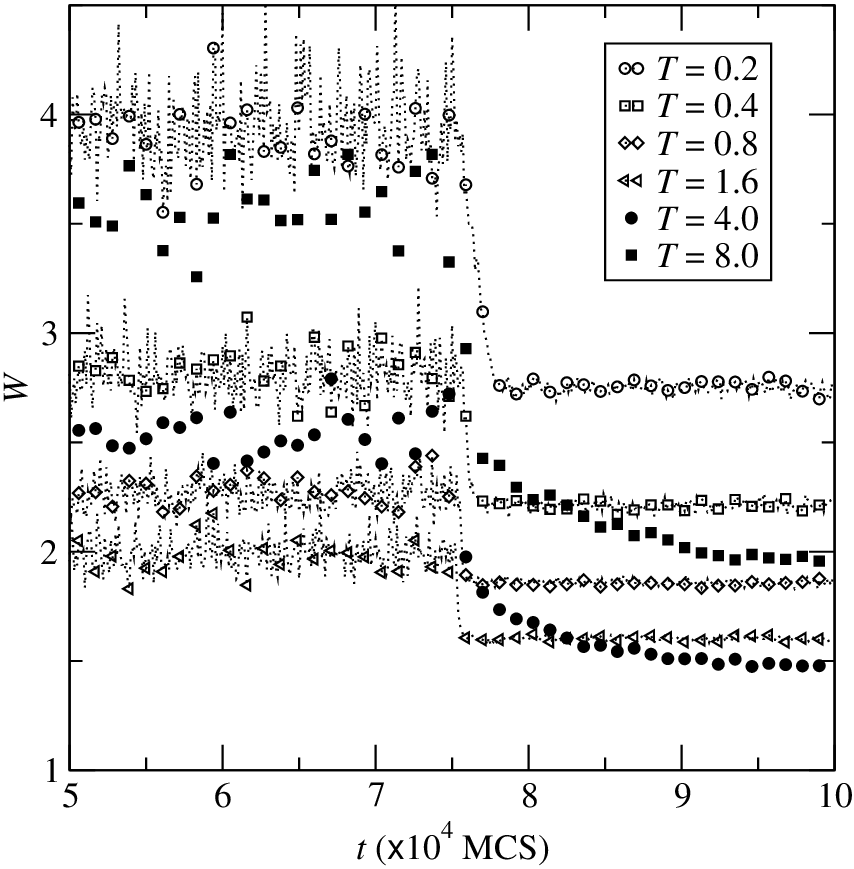}}
\caption{   
Interface width $W$ (in units of the lattice constant) versus time steps $t$ at later times showing surface relaxation. The simulations used KCR segmental
dynamics for different temperatures (in arbitrary units) and 10--20 independent samples.
Other parameters are the same as in Figure\ 1.}
\end{figure}

\begin{figure}
\centerline{\includegraphics*[width=8.5cm]{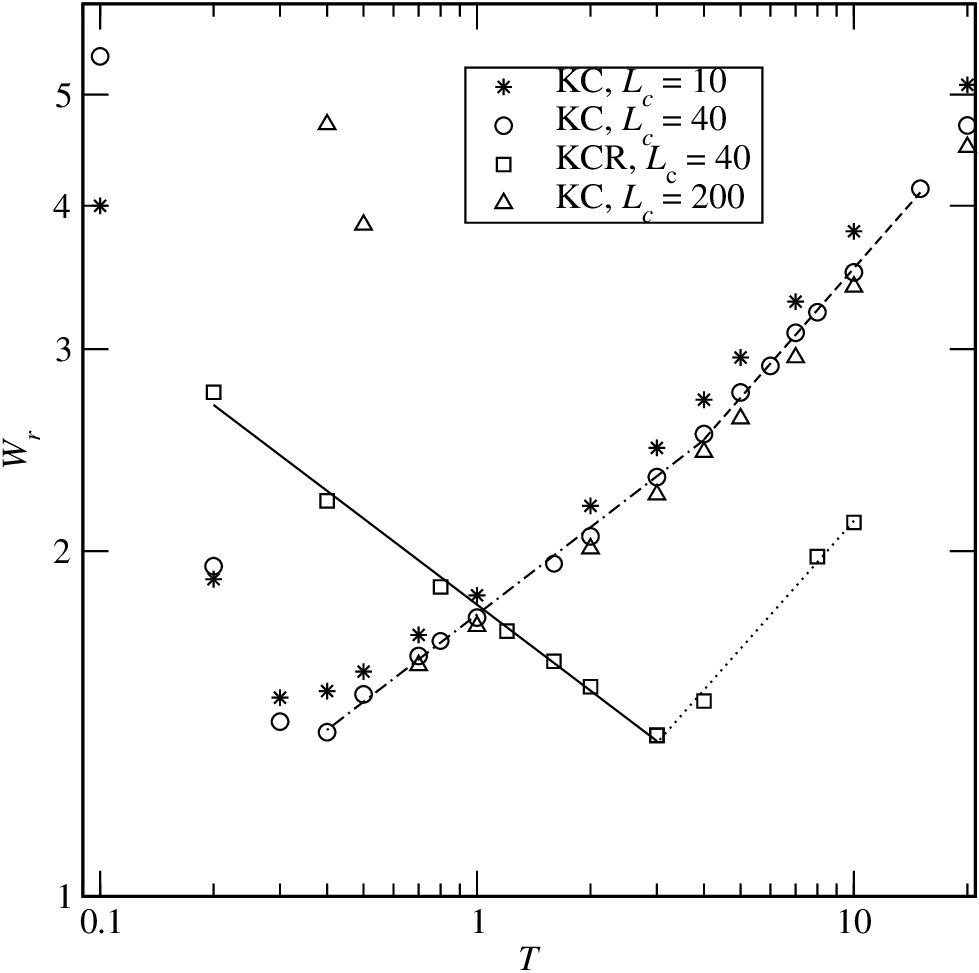}}
\caption{Equilibrium (relaxed) interface width $W_r$ (in units of the
lattice constant) versus 
$T$ using KC, and KCR (squares) 
segmental dynamics with $E=0.5$. Different chain lengths are used with 
the KC dynamics. Other parameters are the same as in Figure\ 1. The 
solid line shows the power-law decay (eq~\ref{eq:decay}) with $\nu=-0.25$ The dotted, dashed, and dotted-dashed lines show the power-law growths (eq~\ref{eq:power}) with $\delta=$0.37, 0.38, and 0.25, respectively. Temperature $T$ and $E$ are in arbitrary units.}
\end{figure}

\begin{figure}
\centerline{\includegraphics*[width=8.25cm]{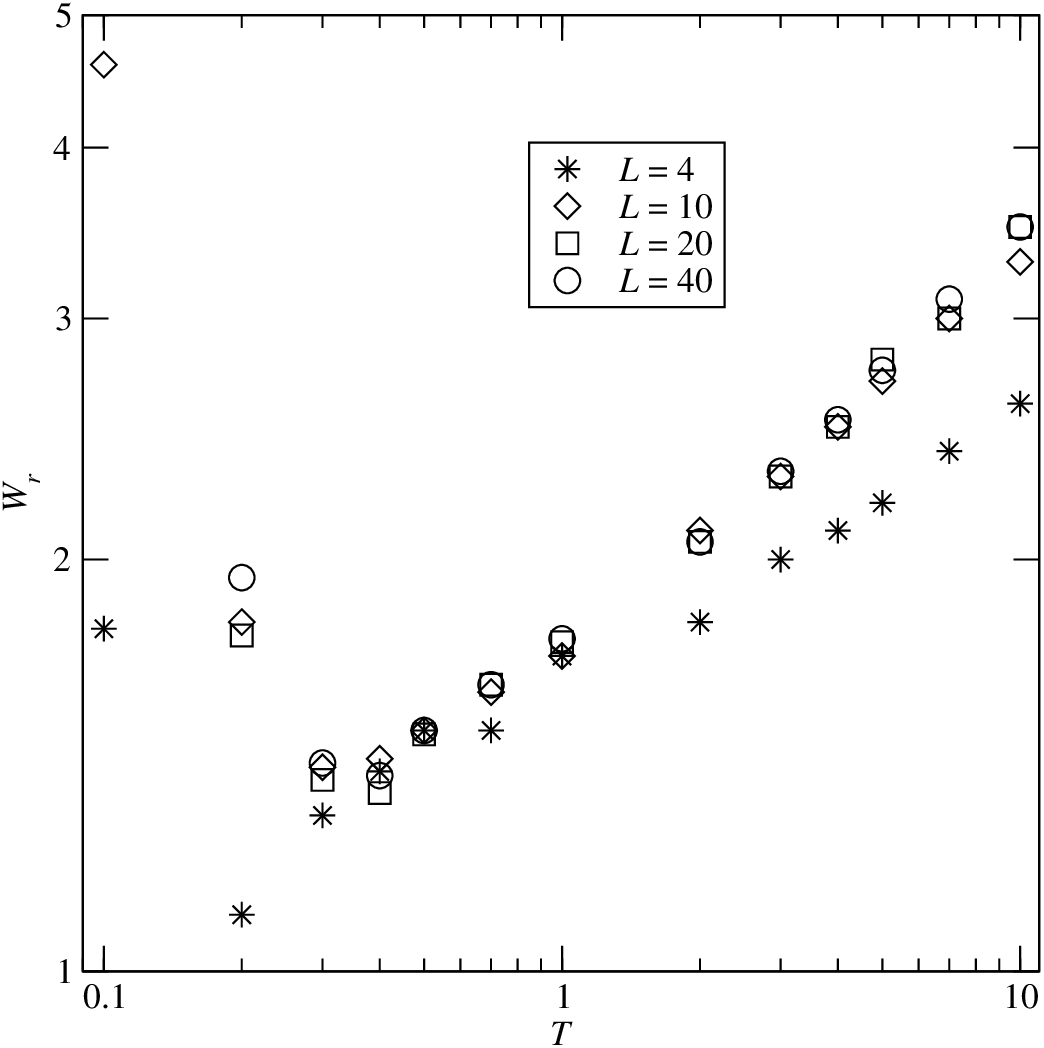}}
\caption{Equilibrium (relaxed) interface width $W_r$ (in units of the
lattice constant) versus 
$T$ using a variety of sample sizes ($L = 4$--40) with KC (circles) segmental dynamics and $E=0.5$. Other parameters are the same as in
Figure\ 1. Temperature $T$ and $E$ are in arbitrary units.}
\label{fig:finite_size}
\end{figure}

The interface width is defined by 
$$W^2 = \frac{\sum_i (h_i-\overline h_i)^2}{L^2},$$
where $h_i$ is the distance in units of the lattice constant from the substrate lattice site $i$ to the deposited polymer farthest from the substrate (in the same $y$-$z$ column) and $\overline h_i$ is the average distance from the substrate. Variation of the interface width with time (in Monte Carlo Steps) for KCR dynamics is
presented in Figure\ 2. We see that the initial power-law growth of the
interface width ($W \propto t^{\beta}$) is followed by a steady-state
value after $10^4$ MCS. After $7.5 \times 10^4$ time steps, the
steady-state interface width $W_s$ is allowed to relax to its equilibrium value
$W_r$ (by ceasing to release new polymer chains into the system) as in Figure\ 3. Results are qualitatively similar for the KC segmental dynamics. Note that the relaxation to equilibrium occurs rapidly except at
high temperatures ($T = 4.0,$ $8.0$). Figure 4 shows the
variation of the equilibrium interface width $W_r$ with the temperature
on a log-log scale using KC and KCR segmental
dynamics. For KC dynamics in the high temperature
regime ($T \ge 4.0$), $W_r$ shows a power-law scaling, 
\begin{equation}
W_r \propto T^{\nu}, 
\label{eq:decay}
\end{equation}
 where $\nu \sim 0.4$. At low temperatures, on the other hand,
there is a tendency for $W_r$ to decay with temperature. With KCR
dynamics, we see a power-law decay, 
\begin{equation}
W_r \propto T^{-\delta},
\label{eq:power}
\end{equation}
 where
$\delta \sim 1/4$. Thus, there is a power-law scaling for the roughening
of the equilibrium width $W_r$ at high temperature and deroughening at
low temperatures. For $L_c=40$ the roughening-deroughening transition temperature for KC dynamics ($T_t\approx 0.35$) is lower than for KCR dynamics,
where $T_t\approx 3.0$. On the other hand, the reduced mobility of longer chains results in an increase in the transition temperature so that $T_t\approx 1.8$ for $L_c=200$ while $T_t\approx 0.32$ for $L_c=10$. To investigate finite-size effects we vary the sample size $L=4$--40 and plot the temperature dependence of the interface width in Figure~\ref{fig:finite_size}. For moderate temperature ($T=$0.4--1.0), no significant finite-size effects are observed. At higher temperatures, thermal mobility increases the lateral height-to-height correlation length, which limits the interface width for small sample sizes ($L=4$). The transverse ($y,z$) components of the polymer chains' radius of gyration increases at low temperatures\cite{bentrem04}, which leads to definite finite-size effects. The transition temperatures are summarized in Table~\ref{tab:transition}. 

\begin{table}

\caption{Transition Temperatures $T_t$ at $E=0.5$}

\label{tab:transition}

\centerline{
\begin{tabular*}{8.25cm}{d@{\extracolsep{\fill}}dcd}
\hline
L_c & L & dynamics & T_t\\
\hline
40 & 4 & KC & 0.20\\
40 & 10 & KC & 0.34\\
40 & 20 & KC & 0.36\\
40 & 40 & KC & 0.35\\
10 & 40 & KC & 0.32\\
200 & 40 & KC & 0.7\\
40 & 40 & KCR & 3.0\\
\end{tabular*}
}

\end{table}

\begin{figure}
\centerline{\includegraphics*[width=8.25cm]{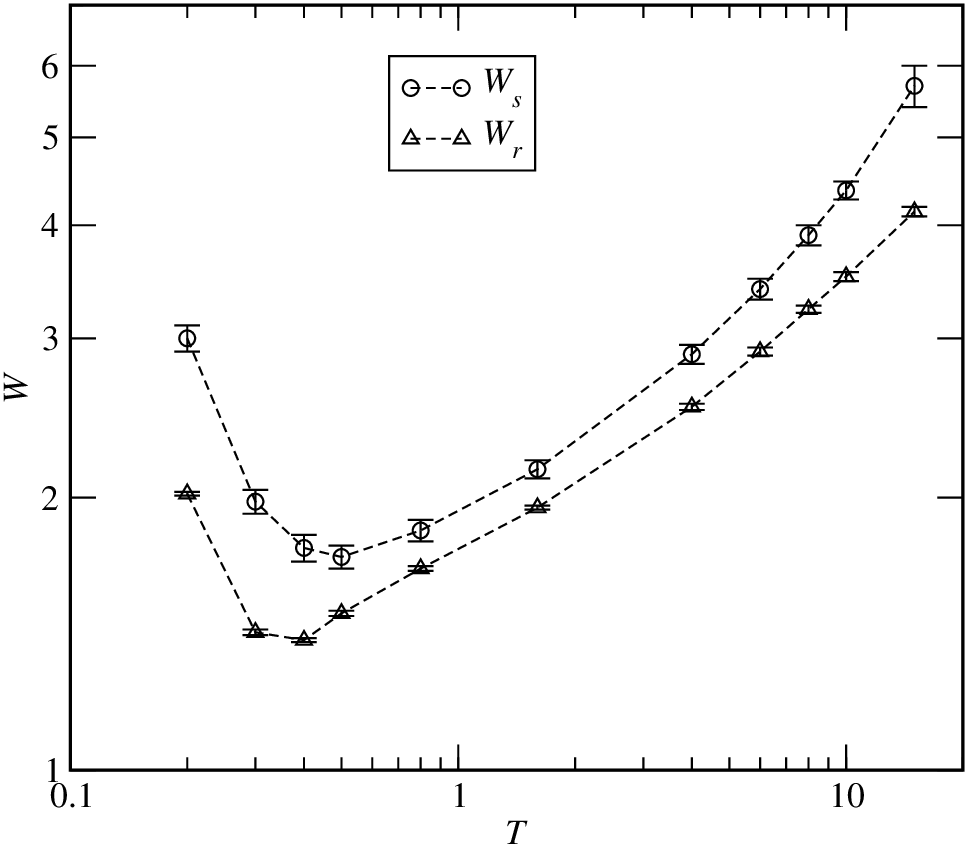}}
\caption{Interface width ($W_s$ and $W_r$, in units of the lattice constant) 
versus $T$ (in arbitrary units) with KC segmental
dynamics. The parameters are the same as in Figure\ 1.}
\label{fig:ws_wr}
\end{figure}

It is also of interest to compare the temperature dependence of the
steady-state ($W_s$) and
equilibrium ($W_r$) interface widths. Figure~\ref{fig:ws_wr} shows their variation with
temperature for KC dynamics. We see that the nonmonotonic dependence
of $W_s$ on the temperature is qualitatively similar to that of $W_r$
with the faster KCR dynamics. Scaling of $W_s$ with $T$ is perhaps more complex
than a single power-law as for $W_r$.

Even for simpler
particle surfaces there exists a complex temperature dependence for roughness with
several mechanisms \cite{stoldt00}, yet the
thermal scaling observed in our polymer interface simulations is
even more complex due to the many degrees of freedom for the chains. At low temperatures we can expect an increase in temperature to
facilitate lateral (intralayer) diffusion \cite{stoldt00} with
relatively less constraint due to entanglement. This will
allow the polymer chains to reach a lower energy configuration
(smoother interface). At temperatures above the
roughening-deroughening temperature, azimuthal (interlayer) diffusion
dominates, and constraint due to entanglement becomes more pronounced creating a (rougher) dynamic interface.

\section{Conclusion}
In summary, the relaxation of the polymer chains, density, and interface 
width depend on the segmental dynamics as well as the temperature. The combination
of slow-to-fast segmental dynamics (KCR) seems to rather effectively equilibrate the system
in our observation time. Simulation with the slower KC dynamics at very low temperatures, while not very efficient for reaching equilibrium, provides insight into the nature of relaxation and the effects of clogging. However, we do not know which segmental dynamics is more 
appropriate for comparison with any given laboratory measurements, though we are able to make some
important scaling predictions: the equilibrium interface width $W_r$ shows 
roughening with the temperature, $W_r \propto T^{0.4}$ in the high temperature
regime and deroughening $W_r \propto T^{-1/4}$ at low temperatures (at
least with KCR segmental dynamics). The transition temperature $T_t$ that separates the roughening and deroughening regimes depends on the polymer chain length and segmental dynamics. The steady-state interface width ($W_s$)
also exhibits a nonmonotonic dependence on temperature, i.e., deroughening
at low temperatures followed by roughening at high temperatures with both KC and KCR dynamics.  

{\bf Acknowledgment.} We would like to thank Jun Xie for assistance with the computer simulations. This work was supported in part by DOE-EPSCoR and NSF-EPSCoR
(\# EPS-0132618) grants. Also, this work was sponsored under Program Element 0603704N by the Oceanographer of the Navy via SPAWAR PMW 150, Captian Bob Clark, Program Manager. The PMW point of contact is Edward Mozley. Participation with the National Science Foundation MRSEC program under the award number DMR-0213883 is acknowledged by R.~B. Pandey.

\end{document}